\documentstyle[12pt,epsfig,rotating]{article}

\newcommand{\lsim}{\raisebox{-0.5mm}{$\stackrel{<}{\scriptstyle{\sim}}$}}

\begin{document}
\begin{flushright}
Bartol Research Institute\\
BA--97--43
\end{flushright}

\vspace*{3cm}
\begin{center}
{\Large\bf Diffraction in Two-Photon Collisions at TESLA}\\[0,5cm]

\vspace*{1cm}

A.\ De Roeck\footnote{e-mail: deroeck@mail.desy.de},\\
{\em Deutsches Elektronen-Synchrotron DESY, D-22603 Hamburg,
Germany,}\\[2mm]
R.\ Engel\footnote{e-mail: rengel@mail.desy.de}\\
{\em Bartol Research Institute, Univ.~of Delaware, Newark
DE 19716, USA}\\[2mm]
and\\[2mm]
A.\ Rostovtsev\footnote{e-mail: rostov@dice2.desy.de}\\
{\em Institute of Theoretical and Experimental Physics ITEP,}\\
{\em Moscow 117259, Russia}

\end{center}

\vspace*{0.5cm}

\begin{abstract}
\noindent
Diffractive reactions have been studied in
hadron-hadron and photon-hadron interactions.
Up to now, such investigations have not been made in
two-photon collisions. In this letter we discuss
the possibility to
measure diffraction dissociation in collisions of real and weakly virtual 
photons at a 500 GeV $e^+e^-$ linear collider. 
\end{abstract}


\section{Introduction}

At a high energy $e^+e^-$ collider 
the interaction of photons with low virtualities 
represents the bulk of the events. 
Similar to hadron interactions, a rise
of the total two-photon cross section with the collision 
energy and a large diffractive contribution are expected. 

Diffractive reactions have been investigated for many years in 
hadron-hadron scattering, photoproduction and, recently, also in
deep-inelastic scattering. They have typically large partial cross
sections and manifest themselves in spectacular topologies with 
rapidity gaps.
Furthermore, new HERA data on $ep$ scattering show that these phenomena
also persist for highly virtual photon-proton interactions, where the photon
probes the short distance structure of the interaction
and perturbative QCD calculations can be applied.
At a high energy linear collider, the photons available for the 
collision can be selected within a broad range of the energy and virtuality.  
This makes two-photon collisions an unique testing ground for
investigations of strong interaction phenomena
\cite{Bartels96a,Brodsky97a}.

In the present work we discuss a possible method of measuring
photon diffraction dissociation in two-photon
collisions at TESLA. In the following, only photons with low virtualities are
considered. However, tagging one or both scattered beam leptons, 
the same method of measurement can be applied to deep-inelastic
$\gamma^\star\gamma$ or $\gamma^\star\gamma^\star$ interactions.

\section{Cross section estimate}

Since soft processes cannot be calculated within perturbative QCD,
it is difficult to obtain a reliable estimate for the expected 
cross section for photon diffraction dissociation.

Simple Regge factorization arguments
imply the following relations for diffraction in $\gamma\gamma$,
$\gamma p$, and $p p$ interactions
\begin{equation}
\sigma_{{\rm SD},\gamma}^{\gamma \gamma} 
\approx
\left(\frac{      b_{{\rm SD},\gamma}^{\gamma p}
\ b_{{\rm SD},p}^{\gamma p}}{
b_{{\rm SD},\gamma}^{\gamma \gamma}\      b_{{\rm SD},p}^{p p}}\right)
\left(\frac{      \sigma_{{\rm SD},\gamma}^{\gamma p}
\ \sigma_{{\rm SD},p}^{\gamma p}}{
      \sigma_{{\rm SD},p}^{p p}}\right)
\hspace*{2cm}
\sigma_{{\rm SD},\gamma}^{\gamma \gamma}
\approx
      \sigma_{{\rm SD},p}^{\gamma p}
\left(\frac{b_{{\rm SD},p}^{\gamma p}}{
b_{{\rm SD},\gamma}^{\gamma \gamma}}\right)
\left(\frac{\sigma_{\rm tot}^{\gamma p}}{
      \sigma_{\rm tot}^{p p}}\right)\ ,
\label{rel-1}
\end{equation}
where $\sigma^{ab}_{{\rm SD},a}$ is the cross sections for 
diffraction dissociation of particle $a$ and $\sigma^{ab}_{\rm tot}$ is
the total cross section in $ab$ collisions. The average slope
characterizing the
diffractive momentum transfer is denoted by $b_{{\rm SD},a}^{ab}$,
respectively.

Both relations (\ref{rel-1}) are equivalent since Regge factorization
also predicts
\begin{equation}
\frac{\sigma_{{\rm SD},\gamma}^{\gamma p}}{
      \sigma_{{\rm SD},p}^{p p}} 
\approx
\left(\frac{b_{{\rm SD},p}^{pp}}{b_{{\rm SD},\gamma}^{\gamma p}}\right)
\left(\frac{\sigma_{\rm tot}^{\gamma p}}{
      \sigma_{\rm tot}^{p p}}\right) \ .
\label{rel-2}
\end{equation}
Assuming a ratio $b_{{\rm SD},p}^{pp}/b_{{\rm SD},\gamma}^{\gamma p}
\approx 7/5$ as indicated by low energy measurements
\cite{Albrow76,Breakstone84a},
Eq.~(\ref{rel-2}) predicts $\sigma_{{\rm SD},\gamma}^{\gamma p}\approx
21$ nb being in good agreement with recent H1 and ZEUS measurements
\cite{Aid95b,Breitweg97a}.
Taking, for example,
$b_{{\rm SD},p}^{\gamma p}/b_{{\rm SD},\gamma}^{\gamma\gamma} = 1$
and an interpolated cross section 
for $p\bar p$ of $\sigma_{{\rm SD},p}^{pp} \approx 4.5 - 5$ mb, 
one gets $\sigma^{\gamma\gamma}_{{\rm
SD},\gamma} \approx 20 - 35$ nb for $\sqrt{s_{\gamma\gamma}} = 200$ GeV
photon-photon collisions. This cross section
refers to one side only, to get the total cross section for diffraction
dissociation it has to be doubled. 

Applying vector dominance arguments, the influence of the photon
virtualities on the diffractive cross section of weakly virtual photons 
can be estimated.
Supposing that photon 1 diffractively dissociates into a system with the
mass $M_{{\rm D},1}$ the differential diffractive cross section
reads
\begin{equation}
\frac{d\sigma^{\gamma\gamma}_{{\rm SD},1}(Q_1^2,Q_2^2)}{dM_{{\rm D},1}}
 \approx
\frac{M_{{\rm D},1}^2}{Q_1^2+M_{{\rm D},1}^2}
\left(\frac{m_\rho^2}{Q_2^2+m_\rho^2}\right)
\left(
\frac{\sigma_{\rm tot}^{\gamma\gamma}(Q_1^2,Q_2^2)}{
\sigma_{\rm tot}^{\gamma\gamma}(0,0)}
\right)
\frac{d\sigma^{\gamma\gamma}_{{\rm SD},1}(0,0)}{dM_{{\rm D},1}}
\label{Qvdm}
\end{equation}
where $m_\rho$, $Q^2_1$ and $Q^2_2$ denote the rho meson mass and the
virtualities of the photons, respectively.

These estimates are subject to large uncertainties. However, models
involving unitarity corrections and non-factorizable contributions
(see for example \cite{Engel95d}) predict a diffractive cross section
compatible with the above mentioned range.

Finally it should be mentioned that a cross section 
of similar size is also expected for quasi-elastic vector meson production 
\cite{Engel95d,Schuler96a}.
Assuming $\sigma_{\rm tot}^{\gamma\gamma} \approx (\sigma_{\rm
tot}^{\gamma p})^2/\sigma_{\rm tot}^{p\bar p}$, about 20 - 30\% of all
$\gamma\gamma$ events are expected to belong to diffraction -- either
quasi-elastic vector meson production or diffraction
dissociation.


\section{Method of measurement}

Experimentally, events with diffraction dissociation can be identified 
using the rapidity gap technique. In non-diffractive 
reactions, rapidity gaps between the final state hadrons are 
exponentially suppressed. In contrast, the differential
cross section $d\sigma_{{\rm SD},\gamma}/d\eta_{\rm gap}$ of diffraction 
dissociation at fixed $\sqrt{s_{\gamma\gamma}}$
is almost independent of the width $\eta_{\rm gap}$ 
of such rapidity gaps.
Hence diffractive particle production can be measured triggering on 
large rapidity gaps. 
\begin{figure}[!hbt]
                 \centerline{
                 \begin{minipage}[t]{14cm}
                 \begin{minipage}[t]{9cm}
                   \unitlength=1cm
                   \begin{picture}(9,6)
                     \put(0,0){\psfig{figure=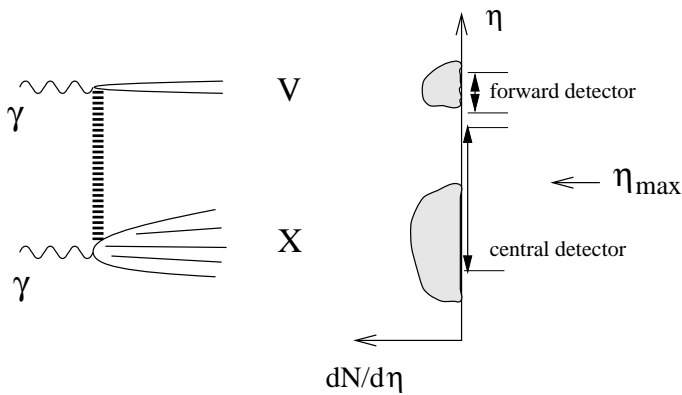,width=9cm}}
                   \end{picture}
                 \end{minipage}
                  \hfill
                 \begin{minipage}[b]{4.0cm}
                   \caption{
Photon single diffraction dissociation and the expected
pseudorapidity distribution of final state hadrons.\label{gg-etamax}}
                   \vspace*{1cm}
                 \end{minipage}
                 \end{minipage}
                 }
\end{figure}
However, the limited angular detector acceptance of the main detector
makes the measurement of both edges of the gap hardly possible. 
As shown by the HERA Collaborations~\cite{Derrick93a,Ahmed94a}, 
the measurement of the 
so-called $\eta_{\rm max}$ distribution can be used instead
to obtain experimental evidence for diffraction.  
The variable $\eta_{\rm max}$ is defined as the pseudorapidity of the
most forward going hadron entering the
central detector part, see Fig.~\ref{gg-etamax}. 
In case of diffraction dissociation, one gets with 
\begin{equation}
\frac{d\sigma_{{\rm SD},\gamma}}{dM^2_{\rm SD}} 
\sim \frac{1}{M^2_{\rm SD}}
\hspace*{1cm}\mbox{and}\hspace*{1cm}
\eta_{\rm max} \sim \ln \left(\frac{M^2_{\rm
SD}}{s_{\gamma\gamma}}\right)
\end{equation}
a differential cross section 
$d\sigma_{{\rm SD},\gamma}/d\eta_{\rm max}$ which is almost independent
of $\eta_{\rm max}$. Events with a small $\eta_{\rm max}$ value
correspond to diffractive final states characterized by a large rapidity gap.

For this analysis, only events with forward
activity (i.e.\ at least 1 GeV energy deposit in the forward tagging
detector) are considered. 
Particles with large pseudorapidities 
could be measured with the planned small angle tagging calorimeter.
It should be emphasized that it is not needed to measure the hadrons in
the entire pseudorapidity range allowed by phase space. The
combination of a forward detector with the central main detector parts is
well suited to find evidence for diffraction (it is also unimportant whether 
there is a gap in the pseudorapidity coverage between these detector parts).
In other words, the variable $\eta_{\rm max}$ used here
measures the pseudorapidity edge of the multi-hadronic system produced
in central main detector.

\begin{figure}[!hbt]
                 \centerline{
                 \begin{minipage}[t]{14cm}
                 \begin{minipage}[t]{9cm}
                   \unitlength=1cm
                   \begin{picture}(9,9)
                     \put(0,0){\psfig{figure=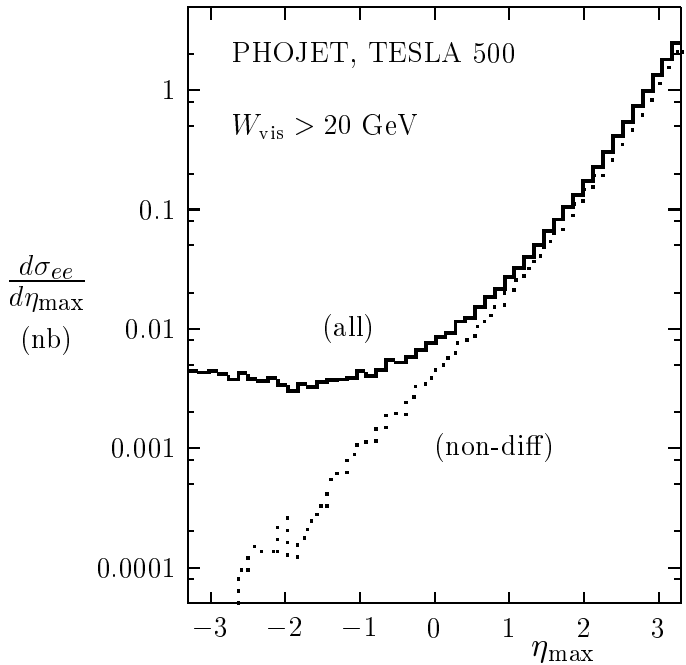,width=9cm}}
                   \end{picture}
                 \end{minipage}
                  \hfill
                 \begin{minipage}[b]{4.0cm}
                   \caption{
The $\eta_{\rm max}$ cross section as calculated using the {\sc Phojet}
MC event generator in
two photon collisions with $W_{\rm vis} \ge 20$~GeV (full curve). The
dotted curve shows the results of calculations for
non-diffractive $\gamma\gamma$ interactions.\label{etamax-w20b}}
                   \vspace*{1cm}
                 \end{minipage}
                 \end{minipage}
                 }
\end{figure}
As an example, a prediction of the $\eta_{\rm max}$ 
cross section is shown in Fig.~\ref{etamax-w20b}
for bremsstrahlung photon-photon interactions in $ee$ collisions at TESLA
($\sqrt{s}_{ee} = 500$ GeV).
The calculations were made using the {\sc Phojet} Monte Carlo event 
generator~\cite{Engel95a,Engel95d}.  
The $\eta_{\rm max}$ distribution is obtained using the hadrons 
produced at pseudorapidities in the central range $-3.3 \le \eta \le 3.3$.
Only events having also particles produced in  
very forward direction ($3.5 \le \eta \le 4.0$) are accepted. 
This trigger on forward going particles is important to suppress
the major background to the diffractive signal due to
highly asymmetric $\gamma\gamma$ collisions 
where the $\eta_{\rm max}$ variable would refer to the
edge of the non-diffractively produced
hadronic final state.
Furthermore, a cut on the visible invariant mass $W_{\rm vis}>20$ GeV
was applied. The visible invariant mass $W_{\rm vis}$ is calculated from
all particles entering the central and the forward detector.

\begin{figure}[!hbt]
\centerline{\psfig{figure=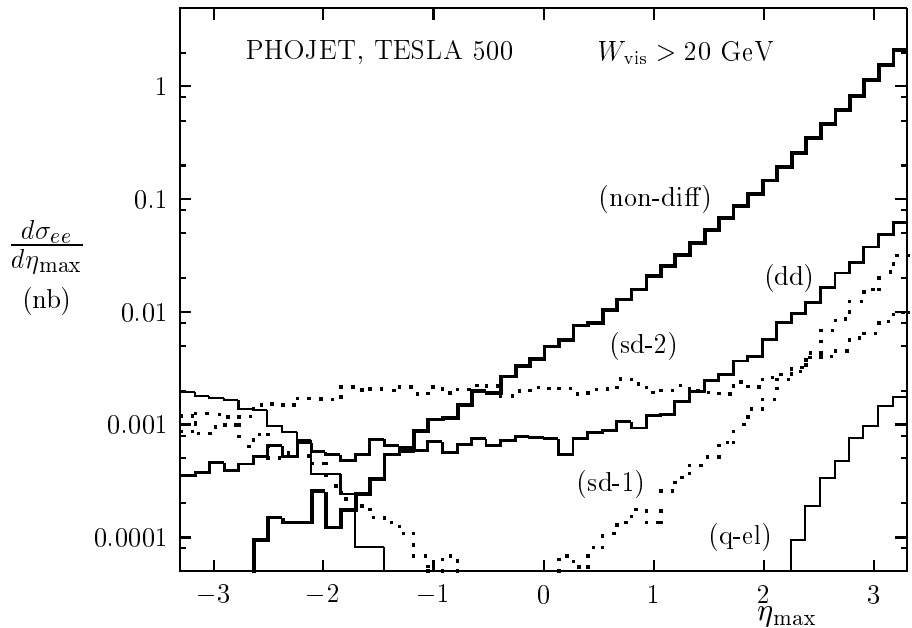,width=12cm}}
\caption{
Breakdown of the  $\eta_{\rm max}$ cross section 
for $W_{\rm vis} \ge 20$~GeV into non-diffractive (non-diff) and
diffractive contributions: quasi-elastic vector meson production (q-el),
single diffraction dissociation of the photon parallel to the $z$ axis
(sd-1), single diffraction dissociation of the photon anti-parallel to the
$z$ axis, and double diffraction dissociation (dd).
\label{etamax-w20a}}
\end{figure}

The exponential suppression of the rapidity gap in non-diffractive events 
is clearly seen (dotted curve). Almost all events with $\eta_{\rm max} < 0$ 
belong to diffraction (diffraction dissociation and quasi-elastic vector
meson production). 
In diffractive events with a large rapidity gap passing the cuts, 
the particles produced in the 
very forward region are mainly decay products of diffractively 
produced vector mesons. 
However, it is not necessary to reconstruct these vector mesons.
In Fig.~\ref{etamax-w20a} the different diffractive contributions are
shown separately for the same kinematics and cuts as used in
Fig.~\ref{etamax-w20b}.
For $W_{\rm vis} > 20$ GeV, the $\eta_{\rm max}$
region from $-2$ to $-1$ is clearly dominated by single diffraction
dissociation of the photon along the $-z$ axis whereas for $\eta_{\rm
max} < -2$ quasi-elastic vector meson production becomes important.
\begin{figure}[!hbt]
                 \centerline{
                 \begin{minipage}[t]{14cm}
                 \begin{minipage}[t]{9cm}
                   \unitlength=1cm
                   \begin{picture}(9,9)
                     \put(0,0){\psfig{figure=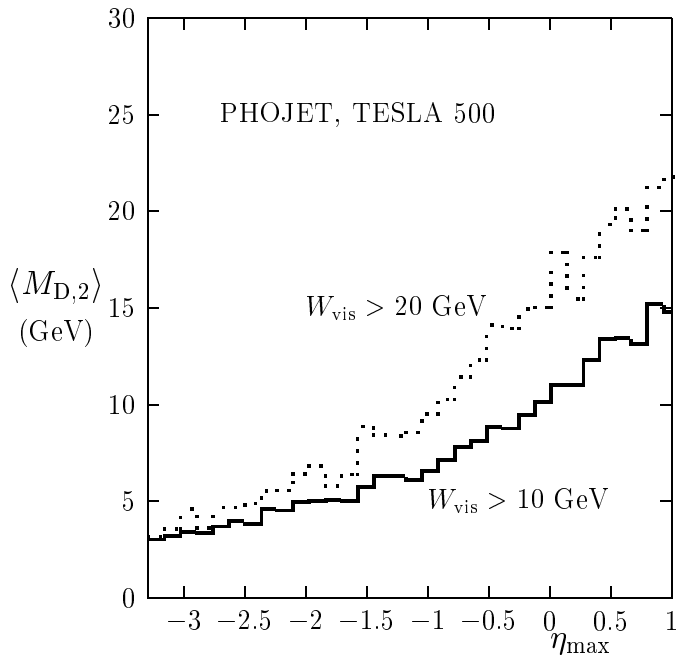,width=9cm}}
                   \end{picture}
                 \end{minipage}
                  \hfill
                 \begin{minipage}[b]{4.0cm}
                   \caption{
Average invariant mass $M_{{\rm D},2}$ of the hadronic system produced by the
dissociation of the photon along the $-z$ direction.
\label{corr-w10-20}}
                   \vspace*{1cm}
                 \end{minipage}
                 \end{minipage}
                 }
\end{figure}
The average mass of the diffractively produced system is shown in
Fig.~\ref{corr-w10-20} for two different $W_{\rm vis}$ cuts.

Figs.~\ref{etamax-w20b},~\ref{etamax-w20a}
have been obtained
without using information on the kinematics of the scattered beam leptons
and, therefore, are dominated by the quasi-real photon interactions.
All final state particles are treated in this analysis in
the same way regardless whether they are scattered beam leptons or
hadrons.   For $W_{\rm vis}\lsim 20$ GeV, almost all of the particles
entering the forward detector are hadrons. 
Lowering the cut on the visible invariant mass increases the
contribution of quasi-elastic vector meson production entering the
$\eta_{\rm max}$ distribution. Furthermore, quasi-elastic vector meson
production events could be identified on an event-by-event basis 
by reconstructing the decay products of the backward scattered vector 
meson.

On the other hand, 
in order to consider large diffractive masses (diffractive 
jet production etc.), one has to
increase significantly the $W_{\rm vis}$ cut. 
For large values of $W_{\rm vis}$, the cross section for having an
electron scattered into the forward detector
becomes comparable with the cross section for finding hadrons there.
Consquently, the scattered beam leptons
have to be removed from the particles used for the trigger in the forward
detector system in order to suppress the non-diffractive background.

In general, the kinematics of the diffractive process (momentum
transfer and invariant diffractive mass) cannot directly be measured.
Tagging the scattered beam leptons will allow one to
reconstruct the kinematics of the diffractive process, however, the
cross section will be suppressed by the photon virtualities.
With 250 GeV beam energy a tagging angle of 40 mrad corresponds to a
photon virtuality of about 50 -- 100 GeV$^2$. This virtuality is too large to
apply simple vector dominance arguments (see Eq.~(\ref{Qvdm}))
to estimate the two-photon 
cross section for diffraction in single- or double-tag events.


\section{Conclusions}

The current study has shown that photon single diffraction dissociation
as well as quasi-elastic vector meson production can be measured in
two-photon collisions at a $\sqrt{s}=500$ GeV $e^+e^-$ linear collider. 
Such measurements
will allow us to check frequently applied Regge factorization 
arguments and will
help us to understand soft multiparticle production. Furthermore, the
determination of the diffractive contribution to the total
$\gamma\gamma$ cross section can be employed to reduce the systematic
uncertainties of many model-based theoretical predictions.

Whereas diffractive and non-diffractive events
can be clearly separated on the basis of the $\eta_{\rm max}$
distribution, only in a restricted phase space
region is it possible to distinguish on a 
event-by-event basis between different diffractive processes such as
vector meson production or single diffraction dissociation.
For
$W_{\gamma\gamma} \lsim  20$ GeV the measurement of quasi-elastic vector
meson production might be possible by reconstructing the meson from the
decay products.

\vspace*{5mm}
\noindent
{\bf Acknowledgments:}\\
One of the authors (R.E.) is grateful to T.\ Gaisser for useful comments.
The work of R.E.\ is supported in part by the U.S.\ Department of Energy 
under Grant DE-FG02-91ER40626.




\end{document}